\documentclass{ieeeaccess}
\usepackage{amsmath,amssymb,amsfonts}
\usepackage{algorithmic}
\usepackage{graphicx}
\usepackage{textcomp}
\usepackage[font=small]{caption}
\usepackage[font=footnotesize]{caption}
\usepackage{multicol}
\usepackage{amsmath,amssymb,subcaption,amsthm,graphicx,stfloats,balance}
\usepackage[backend=bibtex,sorting=none]{biblatex}

\renewbibmacro{in:}{}
\bibliography{BiblioIEEEAccess1}

\def\BibTeX{{\rm B\kern-.05em{\sc i\kern-.025em b}\kern-.08em
    T\kern-.1667em\lower.7ex\hbox{E}\kern-.125emX}}

\begin{document}
	\balance
\history{Date of publication xxxx 00, 0000, date of current version xxxx 00, 0000.}
\doi{10.1109/ACCESS.2019.2925680.DOI}

\title{OAM Mode Selection and Space-Time Coding for Atmospheric Turbulence Mitigation in FSO Communication}
\author{\uppercase{El-Mehdi Amhoud}\authorrefmark{1},  \IEEEmembership{Member, IEEE},
\uppercase{Abderrahmen Trichili} \authorrefmark{1}, (Member, IEEE), \uppercase{Boon S. Ooi} \authorrefmark{1}, (Senior Member, IEEE), and \uppercase{Mohamed-Slim Alouini}\authorrefmark{1},
\IEEEmembership{Fellow, IEEE}}
\address[1]{King Abdullah University of Science and Technology (KAUST), Thuwal 23955-6900, Kingdom of Saudi Arabia }
\tfootnote{This work was partially supported by KAUST-KSU Special Initiative (KKI2) Program, REP/1/3803-01-01.}
\markboth
{Amhoud \headeretal: OAM Mode Selection and Space-Time Coding for Atmospheric Turbulence Mitigation}
{Amhoud \headeretal: OAM Mode Selection and Space-Time Coding for Atmospheric Turbulence Mitigation}

\corresp{Corresponding author: El-Mehdi Amhoud (e-mail: mehdi.amhoud@kaust.edu.sa).}

\begin{abstract}
Orbital angular momentum (OAM) multiplexing has recently received  considerable interest in free space optical (FSO) communications. Propagating OAM modes through free space may be subject to atmospheric turbulence (AT) distortions that cause intermodal crosstalk and power disparities between OAM modes. In this article, we are interested in  multiple-input multiple-output (MIMO)  coherent FSO communication systems using OAM. We propose a selection criterion for OAM modes to minimize the impact of AT. To further improve the obtained performance, we propose a space-time (ST) coding scheme at the transmitter. Through numerical simulations of the error probability, we show that the penalty from AT is completely absorbed for the weak AT regime and considerable coding gains are obtained in the strong AT regime.
\end{abstract}


\titlepgskip=-15pt

\maketitle

\section{Introduction}
In analogy to mode division multiplexing (MDM) in optical fibers where several spatial modes are used for multiplexing, orbital angular momentum (OAM) multiplexing is proposed as a versatile technique to transmit multiple signals  over  free space channels \cite{gibson2004free,Trichili}.  OAM modes are orthogonal which makes them  suitable to co-propagate and carry independent data streams in free space. Laboratory demonstrations have shown beyond 1 Pbit/s free space transmission with a  spectral efficiency exceeding 100 bit/s/Hz using 26 modes \cite{1Pbits}. However, in real-life communication scenarios, OAM beams are subject to atmospheric turbulence (AT) in which the refractive index of the air experiences spatial variations.  The propagation of  OAM beams in the turbulent atmosphere leads to phase-front distortions as well as beam spread and wandering. Moreover, the power of a signal carried by a particular OAM mode is spread to other modes which results in intermodal crosstalk. Furthermore, different OAM modes suffer from different channel losses known as mode-dependent loss (MDL) that causes performance degradation at the system level \cite{TurbulenceEffects}.
\newline
The mitigation of the effects of atmospheric turbulence can be done either at the beam level using adaptive optics (AO) compensation, or by using digital signal processing (DSP) techniques such as  channel coding or equalization. AO aims at correcting the deformations of an incoming wavefront by deforming a mirror, or by controlling a liquid crystal array in order to compensate for aberrations. In \cite{li2014evaluation,ren2014adaptive}, a  real-time AO compensation with wavefront sensors was used to  correct the phase distortions. AO can be also used without wavefront sensors as in \cite{qu2016500}. For DSP approaches, 
MIMO equalization associated with heterodyne detection was shown to mitigate turbulence-induced crosstalk for 4-OAM beams carrying 20 Gbit/s QPSK signals in \cite{CodingFSOMIMO}.  Also, pre-channel combining phase patterns allowed to reduce the crosstalk by 18-dB for a 2-OAM modes transmission in \cite{song2019experimental}.
Moreover, coding techniques such as channel coding \cite{qu2016500} and coded modulation \cite{qu2017two} were also proposed to mitigate atmospheric turbulence in OAM FSO transmissions. Furthermore, In \cite{OAMSpaceCoding}, the authors  proposed a half-rate Alamouti space-time (ST) code and a vertical Bell labs layered ST (V-BLAST) code   along with a sub-optimal zero-forcing channel equalization. The previously proposed AO and DSP  solutions were shown to   significantly improve the  bit-error rate (BER) in the presence of AT. Nonetheless, in all the previous studies, MDL was not taken into consideration. Though even after compensation OAM modes still have different  performances. 
\newline
In the present contribution, we focus on  the effect of  MDL caused by atmospheric turbulence. We show that the amount of MDL depends on the OAM modes considered for multiplexing. Therefore, the BER can be improved by selecting the OAM modes that minimize the MDL.   At the receiver, a maximum likelihood (ML) detection is used  for optimal decoding performance. The proposed selection method completely absorbs the signal to noise ratio (SNR) penalty in weak turbulence conditions. For the strong turbulence regime, and in order to  further enhance the obtained performance,  full-rate full-diversity ST coding at the transmitter is proposed and the performance is evaluated for higher MIMO dimensions. The ST coding scheme brings more than 2.4 dB gain.
\newline
This article is organized as follows: In section II,  spatial multiplexing using OAM in free space transmission and the  atmospheric turbulence effect are described. In Section III, the OAM MIMO communication system model is presented. In Section IV, the mode selection strategy is proposed, and the achieved error probability performance is presented. In Section V, a space-time coding scheme is added at the transmitter to further enhance the obtained performance.  In Section VI, we conclude and set forth future work perspectives.

\section{Orbital Angular Momentum Multiplexing }
\begin{figure*}[!h]
	\centering
	\includegraphics[width=\textwidth,height=4cm]{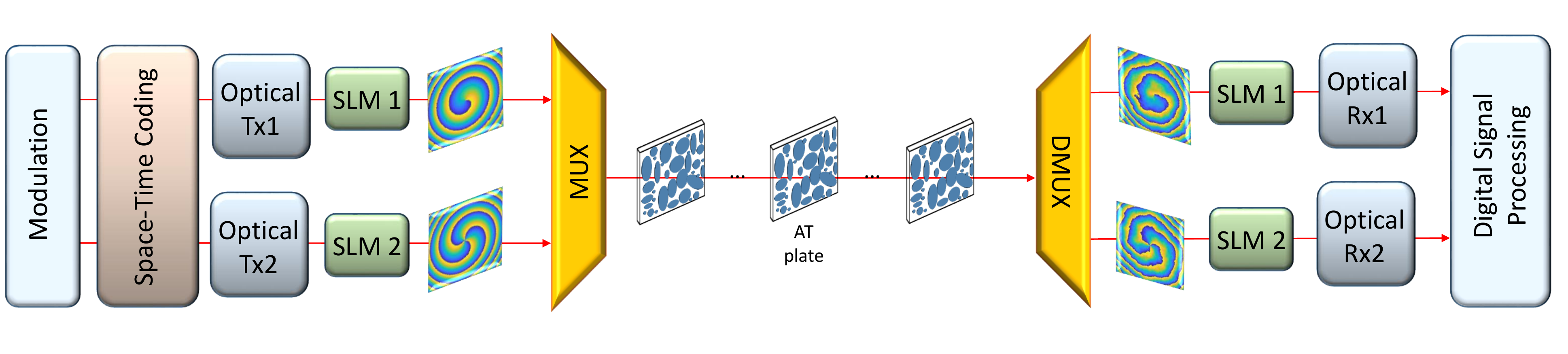}
	\caption{OAM FSO transmission system: SLM: Spatial light modulator, Tx: Transmitter, Rx: Receiver, MUX/DMUX: Multiplexer/De-multiplexer..}
	\label{figMIMO}
\end{figure*}

A lightwave carrying an OAM of $m\hbar$  is a wave having a helical phase-front induced by an azimuthally varying phase term $\exp(im\phi)$, where $m\in \mathbb{Z}$ is the topological charge, $\phi$ is the azimuth and $\hbar$ is the reduced Planck constant. `Orbital' angular momentum should not be confused with `spin' angular momentum which  is related to circular polarization and could have two possible states `right-handed' and `left-handed' circularly polarized. 
To realize OAM multiplexing, single and superpositions of orthogonal beams that have a well-defined vorticity can be used including  Hermite-Gaussian (HG) beams \cite{Ndagano}, Ince-Gaussian beams \cite{bandres2004ince}, Bessel-Gaussian beams \cite{gori1987bessel}, and Laguerre-Gauss (LG) beams \cite{AllenPRA92}. Here, we consider OAM carrying beams derived from LG modes. The field distribution of the LG beams is given by \cite{Doster}:
\begin{multline}
u\left ( r,\phi ,z \right )= \sqrt{\frac{2p!}{\pi(p+\left | m \right |)!}}\frac{1}{w(z)}\left [ \frac{r\sqrt{2}}{w(z)} \right ]^{\left | m \right |}\\
\times L^{\left | m \right |}_p   \left ( \frac{2r^2}{w^2(z)} \right )\times 
\exp\left (\frac{-r^2}{w^2(z)}  \right ) \times 
 \exp\left ( \frac{-ikr^2z}{2(z^2+z^2_R)} \right ) \\ 
 \times \exp\left ( i(2p + \left | m \right | + 1)) 
 \tan^{-1}\left ( z/z_R \right ) \right ) \times \exp\left ( -im\phi \right ),
\end{multline}
where $r$ refers to  the radial distance, $\phi$ is the azimuth angle, $z$ is the propagation distance. \mbox{ $w\left ( z \right )=w_0\sqrt{\left ( 1+\left ( z/z_R \right )^2 \right )}$} is the beam radius at the distance $z$, where $w_0$ is the beam waist of   the Gaussian beam, $z_R=\pi w^2_0/\lambda$ is the Rayleigh range, and $\lambda$ is the optical wavelength carrier \cite{Doster}.  \mbox{$k=2\pi/ \lambda$} is the wavenumber. $ L^m_p(\cdot )$ is the generalized Laguerre polynomial, where $p$ and $m$ represent the radial and angular mode numbers. OAM modes correspond to the subset of LG modes having $p=0$ and $m\neq 0$ which makes OAM multiplexing  not optimal fo achieving the capacity limits of FSO communication systems \cite{zhao2015capacity,chen2016there}. 
\newline\indent
A representation of a $2 \times 2$ MIMO OAM FSO transmission system is shown in Fig. \ref{figMIMO}.
Generation of OAM beams in practice can be realized thanks to different techniques  including spiral phase plates (SPP) \cite{beijersbergen1994helical}, q-plates \cite{marrucci2006optical}, metamaterials \cite{zhao2013metamaterials}, computer-generated holograms (CGHs) loaded on spatial light modulators (SLMs) \cite{heckenberg1992generation}, and  integrated compact devices \cite{cai2012integrated}. An SPP is an optical element having the form of spiral staircases that shapes an incident Gaussian beam into a twisted beam having a helical phase-front. A single SPP allows the generation of a unique OAM mode with a particular topological charge in a stable and efficient manner for a particular wavelength which similar to q-plates that is usually made with liquid crystals with strong wavelength dependence. Metamaterials-based devices and integrated devices are still limited to a low number of OAM modes. However, an SLM can be dynamically addressed to change a digital hologram displayed on a liquid crystal display (LCD) to generate single and superposition of OAM beams in a wide wavelength range from a Gaussian incident beam. SLMs are the most commonly used devices in communication experiments involving OAM beams in different wavelength ranges. At the receiver side, the inverse operation can be performed using the same device to transform an incoming OAM mode back to a Gaussian beam. The idea is to apply an optical scalar product measurement between the incident OAM beam and a CGH with the conjugate phase at the image plane of a Fourier transforming lens \cite{litvin2012azimuthal}. The only inconvenience of using such devices is the diffraction losses at the transmission and reception due to the efficiency of the LCD. The produced Gaussian beam can then be injected into a photodetector to recover the originally encoded signal.
The vorticity of OAM beams propagating in  a FSO media without  atmospheric turbulence is preserved and OAM beams maintain orthogonality as they propagate which can be described by:
\begin{equation}
\int u_p(\boldsymbol{r},z) u_q^\ast (\boldsymbol{r},z)d\boldsymbol{r}=\left\{\begin{matrix}
1~,~~\text{if}~ p=q\\ 
0~,~~\text{if}~ p\neq q
\end{matrix}\right.,
\end{equation}
where $u_q^\ast (\boldsymbol{r},z)$ refers to the normalized field distribution of OAM mode of order $q$ at distance $z$ and $\boldsymbol{r}$ refers to the radial position vector.  
\subsection*{\centering OAM Propagation in Turbulence}
\begin{figure*}[!h]
	\centering
	\includegraphics[width=\textwidth,height=6.5cm]{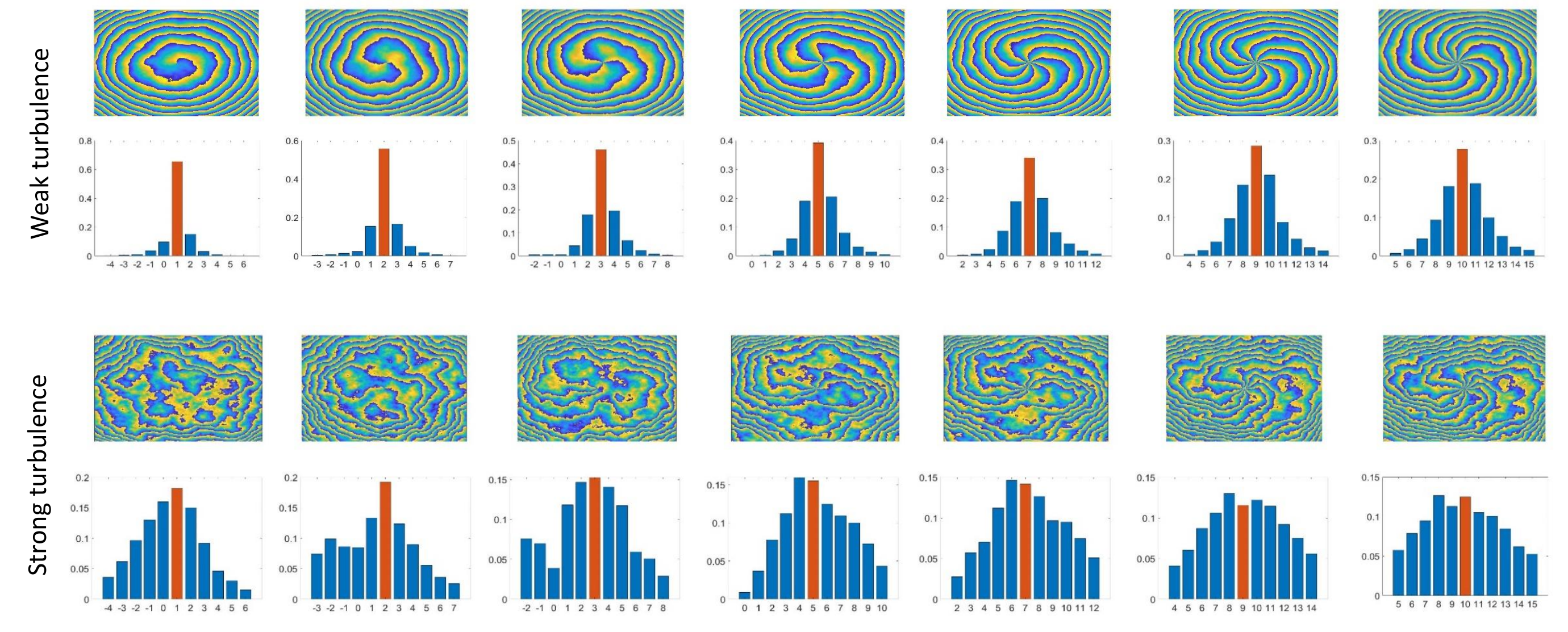}
	\caption{Phase distortion and received power distribution for OAM modes $m=\in \left \{ +1, +2, +3, +5, +7, +9, +10 \right \}$ for weak and strong atmospheric turbulence regimes.}
	\label{figTurbulence_states}
\end{figure*}

The propagation of OAM modes can be affected by atmospheric turbulence induced distortions \cite{TurbulenceEffects,li2014evaluation,CodingFSOMIMO,OAMSpaceCoding}. Atmospheric turbulence is  caused by pressure and temperature fluctuations in the atmosphere which results in a random behavior of the atmospheric refractive index. Due to AT, the power of an initially transmitted OAM mode leaks to other  modes including the ones unused for multiplexing. This phenomenon causes signal overlapping as well as power disparities between OAM modes.
\newline
To emulate atmospheric turbulence, random phase screens are placed along the FSO channel (see Fig. \ref{figMIMO}). These phase screens are generated based on the modified version of the Kolmogorov spectrum given by \cite{andrews1992analytical}:
\begin{eqnarray}
\Phi (\kappa )=0.033C^2_n\frac{\exp(-\kappa ^2/\kappa_l^2)}{(\kappa ^2+1/L_0)^{11/6}}f(\kappa,\kappa_l),
\end{eqnarray}
where \mbox{$f(\kappa,\kappa_l)=[1+1.802(\kappa/\kappa_l)-0.254(\kappa/\kappa_l)^{7/6}]$}. $C^2_n$ is the refractive index structure parameter, $L_0$ is the outer scale of the turbulence, $\kappa_l=\frac{3.3}{l_0}$, with $l_0$ is the inner scale of the turbulence. The  turbulence strength in an FSO channel is given by the Rytov variance defined as \mbox{$\sigma _R^2=1.23C^2_n(2\pi/\lambda )^{7/6}z^{11/6}$},  where $\lambda$ is the carrier wavelength and $z$ is the propagation distance. We note that for $\sigma _R^2<1$ ($\sigma _R^2>1$), the system is operating under a weak (strong) turbulence regime. The beam waist at the transmitter for all beams is set to $\omega_0=1.6$ cm to ensure a minimal beam waist at the receiver plane. On the other hand, we assume that the optical receiver is large enough to collect all received OAM beams. To create the desired OAM modes, SLMs with LCD of dimension \mbox{$512 \times 512$} pixels are used.  The propagation distance is set to $z=1~$km, the inner and outer scales of turbulence are set to $l_0=5~$mm and $L_0=20~$m, respectively. AT is emulated by placing $20$ random phase screens each  $50~$m. Each phase screen is evaluated  as the Fourier transform of a complex random distribution with zero mean and variance equal to  $\left ( \frac{2\pi}{N\Delta x} \right )^2\Phi (\kappa )$, where $N=512$ is the array length and $\Delta x=5~$mm is the grid spacing assumed to be equal in both dimensions $x$ and $y$. Propagation through the turbulent atmosphere is simulated using the commonly used split-step Fourier method at wavelength \mbox{$\lambda=1550$ nm}. For weak (strong) turbulence, we set $C^2_n=10^{-14}$ ($C^2_n=10^{-13}$), respectively. 
\newline
The phase-front of OAM beams with  topological charges $m\in \left \{ +1, +2, +3, +5, +7, +9, +10 \right \}$ and their corresponding mode purity graphs are presented in Fig. \ref{figTurbulence_states} for different atmospheric turbulence regimes after \mbox{1 km} of propagation. For the weak turbulence regime, the phase-front of OAM modes is affected but can still be recognizable. In this case, the spread of optical power to other OAM modes is relatively low (blue bars) and more power is conserved in the originally launched mode (red bars).  However, for a strong turbulence regime, we can clearly see that the phase of OAM beams is severely impacted. This results in a wide spread and a considerable power leakage to other OAM modes. Moreover, we also notice that due to beam spread, OAM modes with higher topological charges suffer from more important power spread (see mode purity graphs in Fig. \ref{figTurbulence_states}) which impacts the OAM capacity and system outage as shown in \cite{Huang}. 
\newline
In addition to  crosstalk, the break of the orthogonality between OAM states due to AT causes transmitted modes to have different losses.  MDL creates an SNR imbalance of the received signals, and since the overall system performance is averaged over all received signals, then the performance is limited by the signal with the most degraded SNR.
MDL was intensively studied for few-mode fibers, and multi-core fibers optical communications \cite{amhoud,Awwad_JLT,Winzer} and was shown to be the main performance limiting factor. In OAM FSO communications, the impact of MDL was only investigated in the case of laterally displaced OAM beams \cite{Ndagano}.  
\newline
In the next section, we show that MDL also reduces the performance of OAM FSO systems. Moreover, the selected OAM modes for multiplexing determines the level of MDL of the MIMO system.  

\section{System Model }
The transmission of $M$  OAM beams in the turbulent FSO channel can be described by an $M \times M$ MIMO system as:
\begin{eqnarray}
\mathbf{y}=\mathbf{Hs}+\mathbf{n},
\label{eq3}
\end{eqnarray}   
where $\mathbf{s}\in \mathbb{R}^M$ is a vector of modulated symbols. $\mathbf{y}\in \mathbb{R}^M$ is the received symbols vector. $\mathbf{n}\in \mathbb{R}^M$ denotes the noise assumed to be additive white Gaussian with variance $N_0$ per complex dimension. The transmission matrix $\mathbf{H}$ is a square matrix of dimension  ${M}$.  It  represents the turbulent FSO channel, where each diagonal element $h_{p,p}$ is the efficiency of channel $p$ which represents the amount of power that remains in OAM state $p$. The crosstalk between OAM states due to turbulence is represented by the channel inputs $h_{p,q}$ and can be expressed as follows:
\begin{equation}
h_{pq}=\int u_p(\boldsymbol{r},z) u_q^\ast (\boldsymbol{r},z)d\boldsymbol{r}.
\label{hij}
\end{equation}
The crosstalk fading $h_{pq}$ was demonstrated in \cite{Funes} to obey a Johnson $S_B$ distribution. However, the latter distribution is analytically  intractable which makes theoretical derivations not feasible.
After propagating through the turbulent atmosphere, symbols carried by OAM beams are subject to transmission errors. To retrieve the original signals, at the receiver side, we can estimate the symbols using a maximum likelihood (ML) decoder.
The latter gives the optimal decoding performance and hence, can be considered as a powerful DSP technique to recover signals after being distorted by AT \cite{gupta2012ber}. The ML criterion estimates the transmitted symbols vector $\mathbf{s}$ by minimizing the following Euclidean distance: 
\begin{equation}
\hat{\mathbf{s}}_{\text{ML}}=\underset{\mathbf{s}\in \mathfrak{C}^M}{\text{argmin}}\left \| \mathbf{y}-\mathbf{Hs} \right \|^2,
\end{equation}
where $\hat{\mathbf{s}}_{\text{ML}}$ is the estimated symbols vector and $\mathfrak{C}$ is  the set of all possible symbols. The ML detection can be realized by an exhaustive search over all combination possibilities in the set $\mathfrak{C}^M$. However, it is often considered not to be feasible in practice due to the exponential computational complexity that scales with $\left | \mathfrak{C} \right |^M$, where $\left | \mathfrak{C} \right |$ is the constellation size. To overcome  this issue, reduced search lattice decoders can also return the ML solution at a reduced complexity. By applying a complex to real transformation to Eq. (\ref{eq3}), the ML decoding metric can be rewritten as:
\begin{equation}
\hat{\mathbf{s}}_{\text{ML}}=\underset{\mathbf{s}_{\mathfrak{R}} \in \mathfrak{C}^{2M}}{\text{argmin}}\left \| \mathbf{y}_{ \mathfrak{R}}-\mathbf{H}_{eq}\mathbf{s}_{\mathfrak{R}} \right \|^2,
\end{equation}

\begin{equation*}
\text{with}~\mathbf{y}_{\mathfrak{R}}=\left [ \text{Re}(\mathbf{y})~ \text{Im}(\mathbf{y}) \right ]^{T},~~\mathbf{s}_{\mathfrak{R}}=\left [ \text{Re}(\mathbf{s})~ \text{Im}(\mathbf{s}) \right ]^{T},
\end{equation*}
\begin{equation*}
\mathbf{H}_{eq}=\begin{bmatrix}
\text{Re}(\mathbf{H}) & -\text{Im}(\mathbf{H})\\ 
\text{Im}(\mathbf{H}) &  \text{Re}(\mathbf{H})
\end{bmatrix}.
\end{equation*}
Consequently, the problem of minimizing the quadratic error becomes a search problem  for the closest point $\mathbf{s}_{\mathfrak{R}}$ in the lattice generated by the equivalent real matrix $\mathbf{H}_{eq}$.
In our work, we consider the sphere decoder (SD) \cite{viterbo1999universal} for decoding the received signals.  The SD searches for the optimal point in a finite sphere centered on the received point  $\mathbf{y}_{\mathfrak{R}}$. Consequently, the complexity is reduced, and more importantly becomes independent from the constellation size and is approximately equal to $M^6$ \cite{damen2000lattice}.
\newline
The error probability of the transmission is given by \cite[chap.~4]{Proakis}:
\begin{equation}
P_{\text{e}}=\sum_{\mathbf{s}_i\in \mathfrak{C}^M}{} \text{P}_\text{r} \left \{ \mathbf{s}_i \right \}\text{P}_\text{r} \left \{ \mathbf{s}_i\neq  \hat{\mathbf{s}}_{i,\text{ML}}  \right \},
\end{equation}
where the sum is over all possible transmitted symbol vectors $\mathbf{s}_i$.
\section{OAM Mode Selection}
\begin{figure*}[!t]
\centering
\begin{subfigure}[b]{0.5\textwidth} 
	\includegraphics[width=\columnwidth,height=7.5cm]{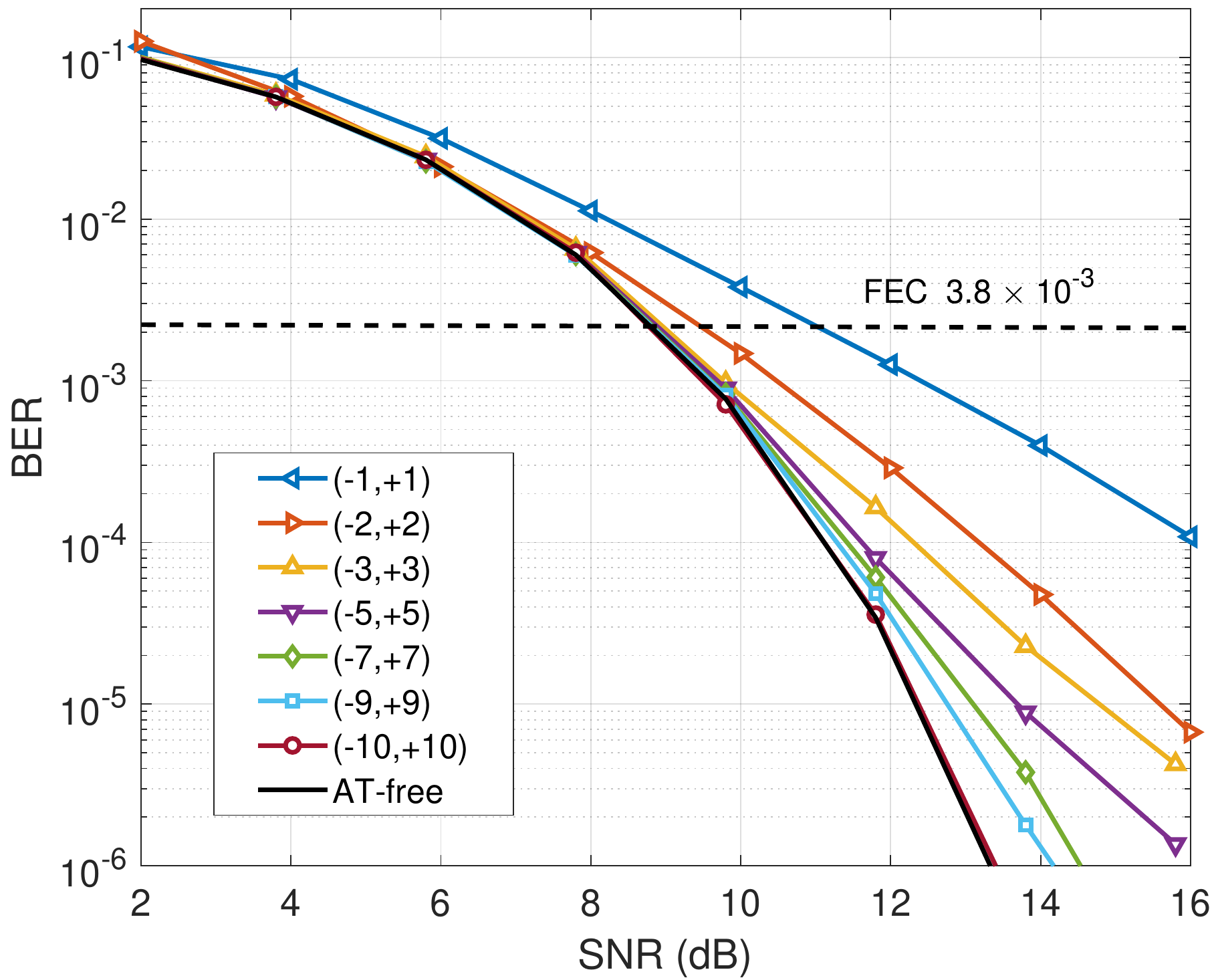}
		\caption{\footnotesize{  Weak turbulence regime, $C^2_n=10^{-14}$. } }
\end{subfigure}~
\begin{subfigure}[b]{0.5\textwidth} 
	\includegraphics[width=\columnwidth,height=7.5cm]{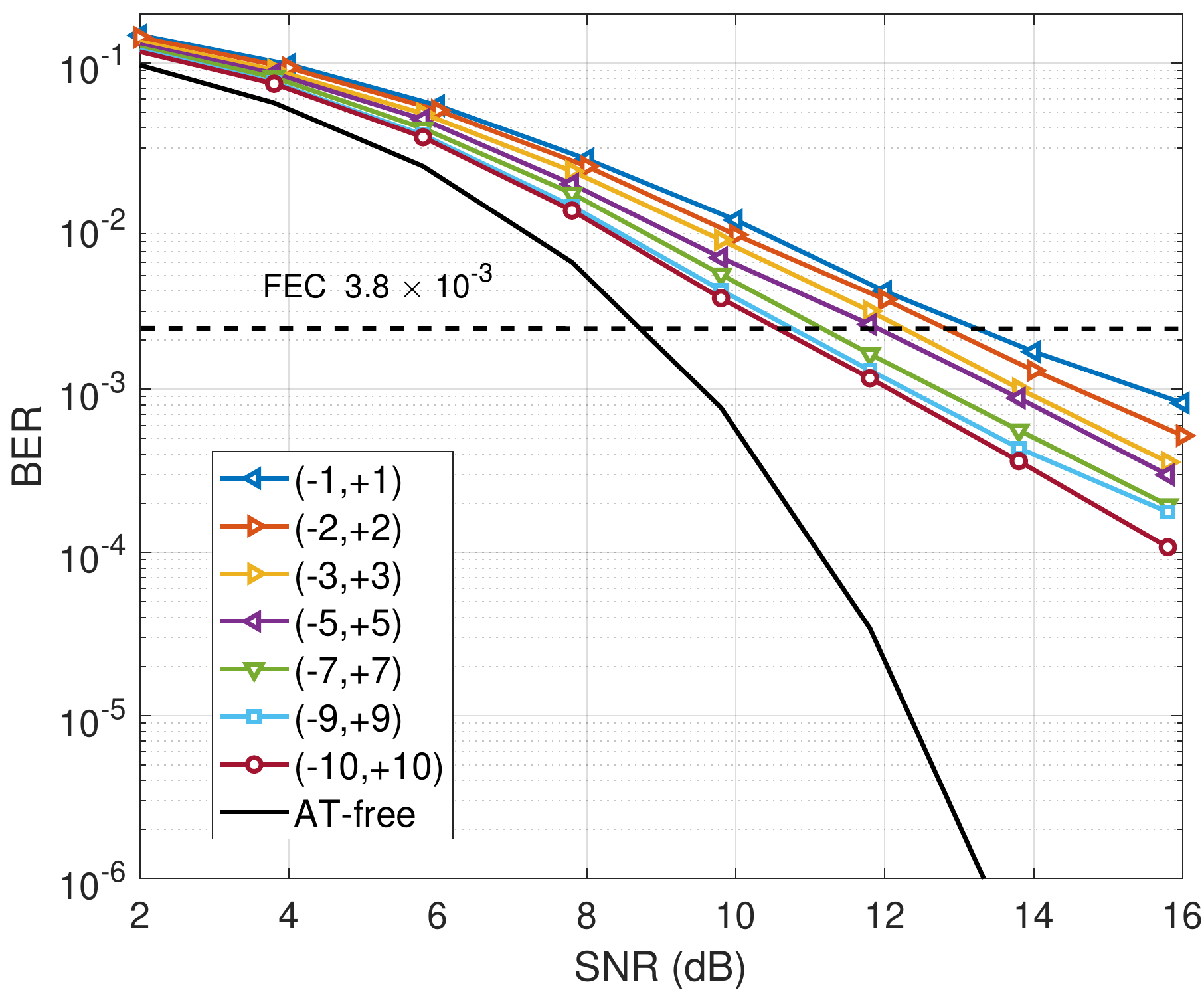}
	\caption{\footnotesize{Strong turbulence regime, $C^2_n=10^{-13}$. } }
\end{subfigure}
\caption{BER versus SNR for different OAM  sets for a $2 \times 2$ MIMO system. (a): Weak turbulence regime, (b): Strong turbulence regime.}
	\label{noncoded2x2}
\end{figure*}

\begin{figure*}[!ht]
	\centering
	\begin{subfigure}[b]{0.5\textwidth} 
		\includegraphics[width=\columnwidth,height=7.5cm]{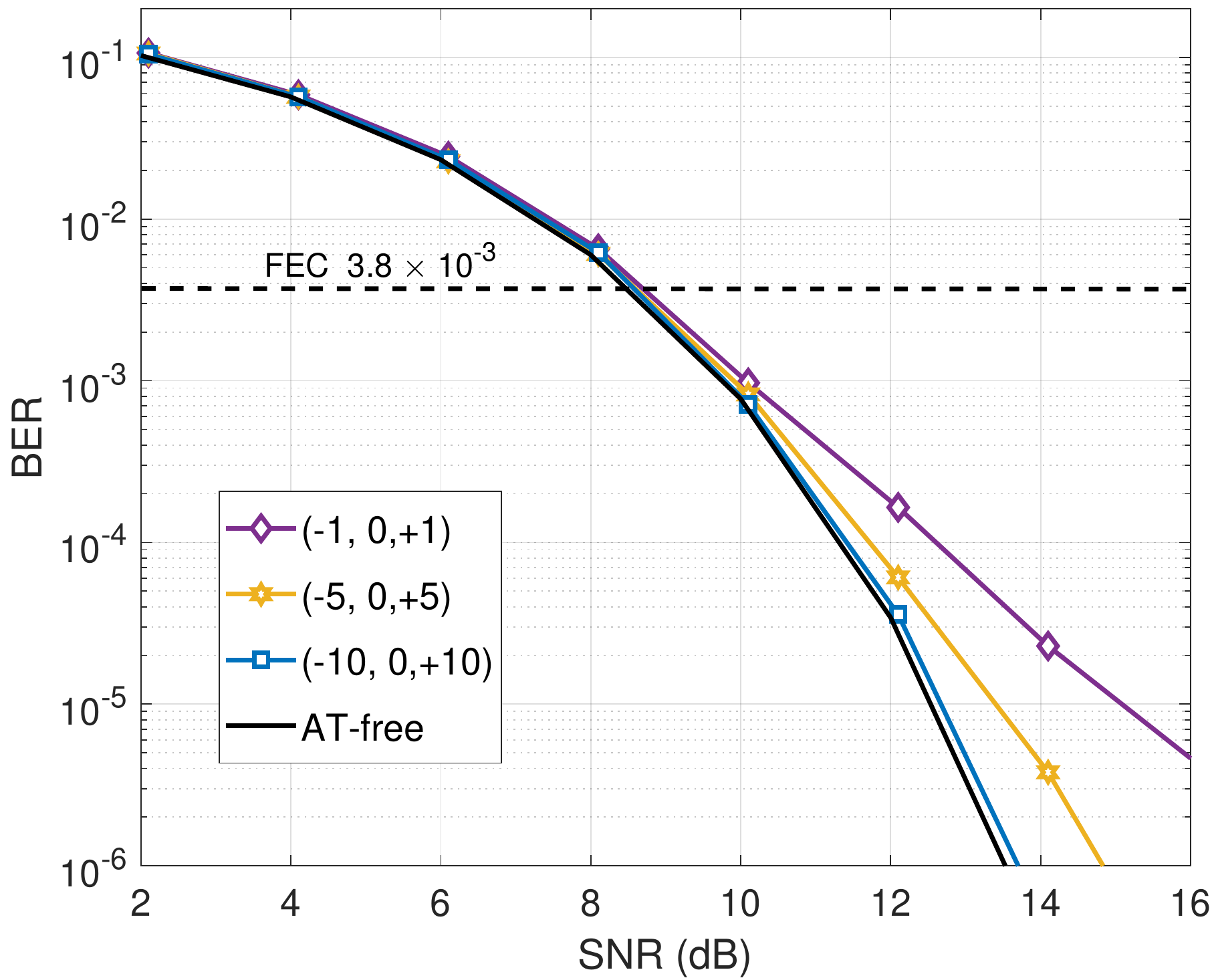}
		\caption{\footnotesize{  Weak turbulence regime, $C^2_n=10^{-14}$. } }
	\end{subfigure}~
	\begin{subfigure}[b]{0.5\textwidth} 
		\includegraphics[width=\columnwidth,height=7.5cm]{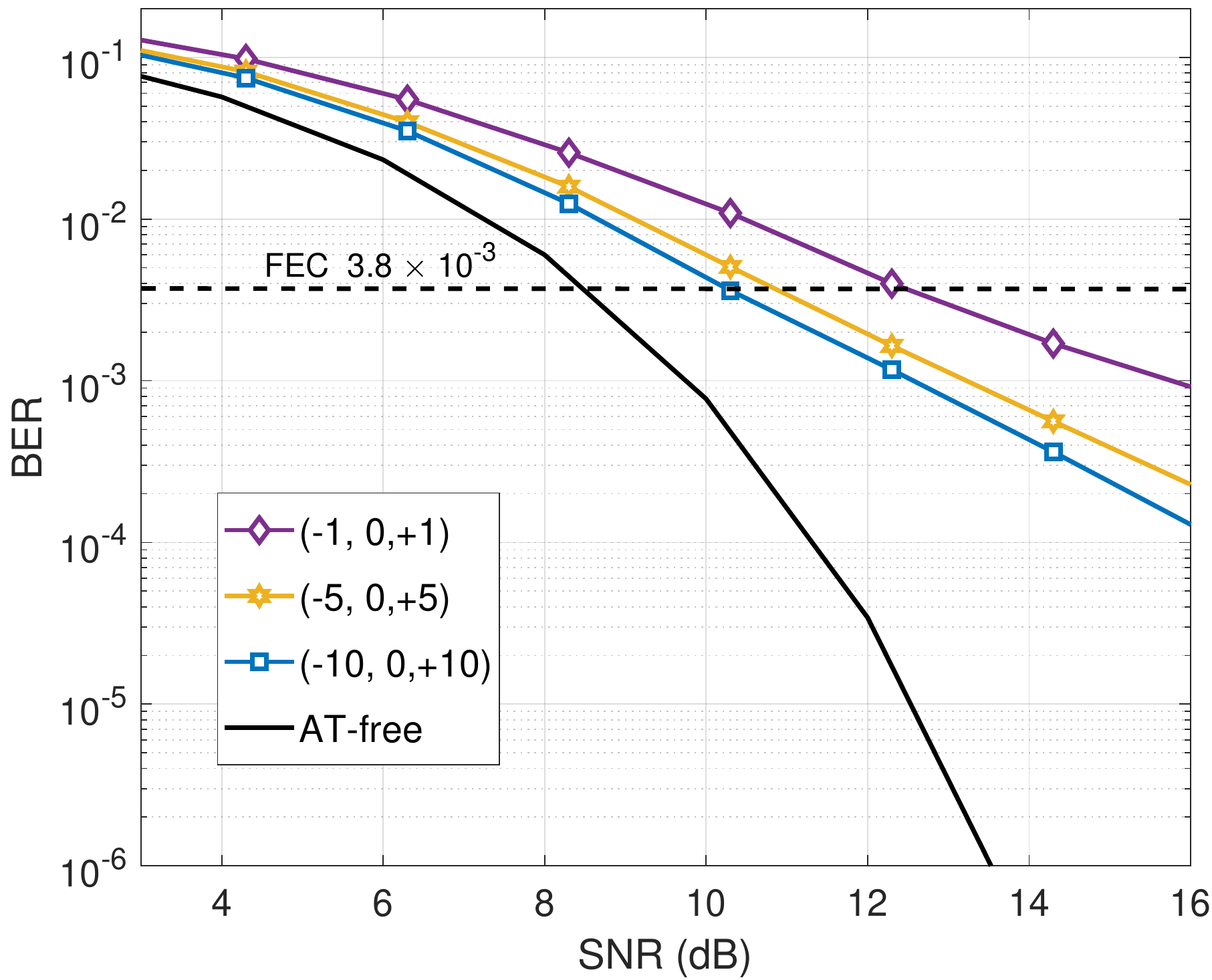}
		\caption{\footnotesize{Strong turbulence regime, $C^2_n=10^{-13}$. } }
	\end{subfigure}
	\caption{BER versus SNR for different OAM  sets for a $3 \times 3$ MIMO system. (a): Weak turbulence regime, (b): Strong turbulence regime.}
	\label{noncoded3x3}
\end{figure*}

Mode-dependent loss  is given by the decibel ratio  of the maximum to the minimum eigenvalues of the channel matrix  $\mathbf{H}$ as \cite{Awwad_JLT,Winzer}:
\begin{equation}
\text{MDL}=10\log_{10}\left ( \frac{\lambda_{\text{max}}}{\lambda_{\text{min}}} \right ),
\end{equation} 
where $\lambda_{\text{max}}$ (resp. $\lambda_{\text{min}}$) is the maximum (resp. minimum) eigenvalue of the channel matrix.
\newline
Let $\mathcal{S}$ be  the set of available OAM modes for multiplexing, we aim to select the set of OAM modes ${\mathcal{S}_{\text{p}}}$ that gives the best performance in terms of the error probability. 
In \cite{Huang}, the authors found the set of OAM modes that maximizes the asymptotic average transmission rate when all OAM modes were considered as independent channels in an IM/DD system. However, maximizing the overall channel capacity does not guarantee the optimal reliability performance of the system. Hence, different  BERs are observed for each OAM mode which impacts the average BER of the system. In this work, a joint detection of all OAM modes based on ML decoding is considered. In the presence of modal crosstalk only and without MDL, the ML decoder is capable of uncoupling all channels and achieves the same BER performance as a crosstalk-free transmission. However, in the presence of MDL, the performance of the ML decoder degrades. Consequently, to obtain the optimal BER performance, we focus on minimizing the MDL of the MIMO channel. 
\begin{figure*}[!h]
	\begin{subfigure}[b]{0.5\textwidth} 
		\centering
		\includegraphics[width=\columnwidth,height=7.5cm]{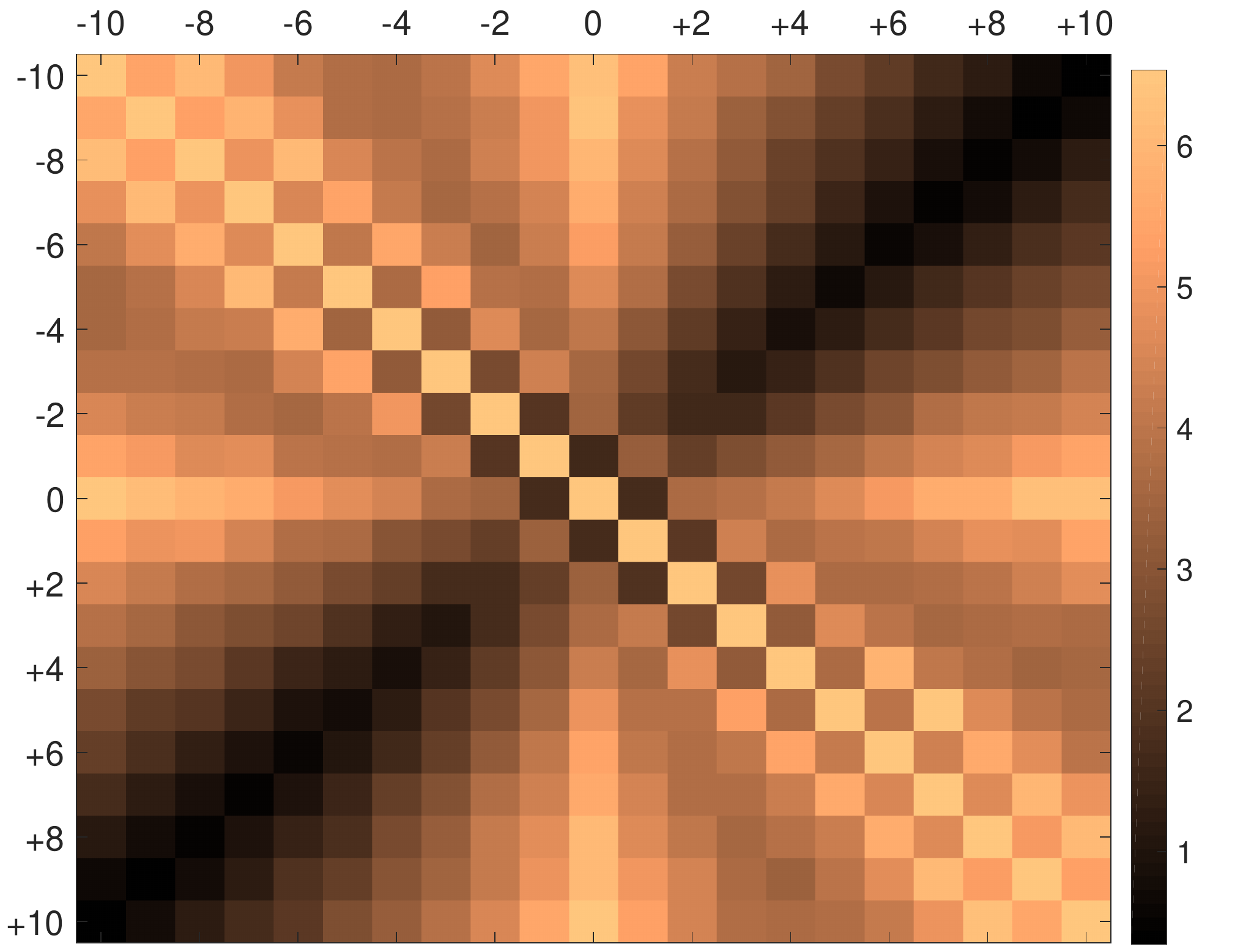}
		\caption{\footnotesize{  Weak turbulence regime, $C^2_n=10^{-14}$. } }
	\end{subfigure}~
	\begin{subfigure}[b]{0.5\textwidth}
		\centering
		\includegraphics[width=\columnwidth,height=7.5cm]{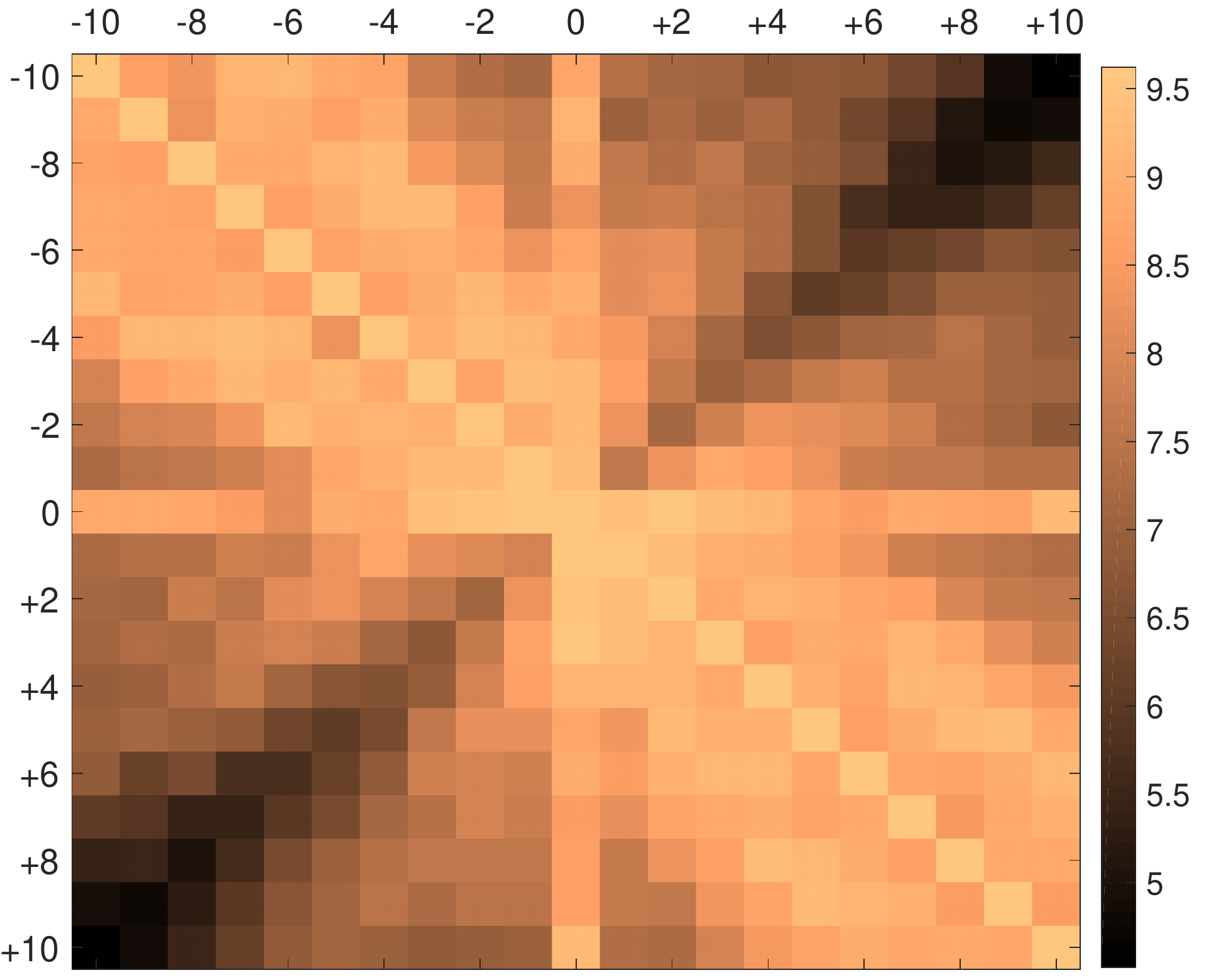}
		\caption{\footnotesize{Strong turbulence regime, $C^2_n=10^{-13}$. } }
	\end{subfigure}
	
	\caption{MDL for different sets of OAM modes $\left ( p, q \right )$. (a): Weak turbulence regime, (b): Strong turbulence regime.}
	\label{MDL}
\end{figure*}
\newline
Moreover, the MDL of the channel does not depend on the SNR but only depends on the power ratio between the maximum and minimum received powers. Hence, we aim to find the optimal set of modes that satisfies the following condition:
\begin{equation}
{\mathcal{S}_{\text{p}}}= \underset{{\mathcal{S}_{\text{sub}}}\subset \mathcal{S} }{\text{argmin}}~\text{MDL},
\label{minimization}
	\end{equation}
where $\mathcal{S}_{\text{sub}}$ spans all the possible subsets of $\mathcal{S}$.
\newline
To have an insight on the MDL levels for different sets of OAM modes, we consider a $2 \times 2$ MIMO transmission. We compute the MDL averaged over $10^{6}$ channel realizations by considering the same simulation parameters of the previous section.
In Fig. \ref{MDL}(a) and \ref{MDL}(b), the  MDL  is shown for all possible combinations of OAM modes with topological charges $p$ and $q$ spanning the set \mbox{$\mathcal{S}=\left \{ -10, -9,...,+9, +10 \right \}$}. For both the weak and strong AT regimes, the minimum values of the MDL were found for OAM modes having opposite topological charges ($p=-q$) (see anti-diagonal elements in Fig. \ref{MDL}). Moreover, for  OAM modes satisfying the previous condition, the MDL decreases as the topological charge increases. Hence, the lowest MDL corresponds to the set $\left ( -10, +10 \right )$. Another important observation is that a pair of modes with a low level of crosstalk does not necessarily achieve a low level of MDL. For example the set  $\left ( -1, +1 \right )$ have more crosstalk than $\left ( -1, +10 \right )$ because OAM mode $-1$ will spread more power to $+1$ than to $+10$ which is much further. Nonetheless, the MDL of the set $\left ( -1, +1 \right )$  is lower than the set $\left ( -1, +10 \right )$.
\newline
Furthermore, we have resolved the minimization problem of Eq. (\ref{minimization}) and found the optimal sets of OAM modes that minimize MDL for  higher MIMO dimensions. In  \mbox{Table \ref{Table}}, the optimal sets are given for different AT strengths given by the values of the refractive index structure parameter. As shown from the table, further OAM modes having opposite topological charges allow obtaining the lowest MDL. 
\begin{table}[!h]
	\caption{Optimal OAM sets for different MIMO systems and AT strengths. }
	\centering
	\begin{tabular}{lll}
		\hline 
		MIMO&Refractive index structure parameter&Optimal OAM set  \\
		\hline 		
		$2 \times 2$ & $C^2_n=10^{-13}$  &  $\left \{ -10, +10 \right \}$     \\
		$3 \times 3$ & $C^2_n=6 \times10^{-14}$  & $\left \{ -10,0, +10 \right \}$      \\
		$4 \times 4$ & $C^2_n=3 \times10^{-14}$  & $\left \{ -10,-5, +5, +10 \right \}$  \\
		$5 \times 5$ & $C^2_n=10^{-14}$  & $\left \{ -10,-5, 0, +5, +10 \right \}$     \\
		\hline
	\end{tabular}
	\label{Table}
\end{table}
\newline
To examine the efficiency of the proposed mode selection approach on the error probability performance, we consider the  transmission of OAM modes in a $2 \times 2$ and $3 \times 3$ MIMO configurations.
\newline
At the transmitter, bits are modulated to form QPSK symbols that are sent on the propagating modes. At the receiver, a sphere decoder is implemented. We measure the BER performance using Monte-Carlo simulations, and for each simulated point, we record a minimum of 100 bit errors.
\newline
In Fig. \ref{noncoded2x2}, we consider a $2 \times 2$ MIMO system, we plot the BER as a function of  SNR  for different sets of OAMs in the weak and strong turbulence regimes. The atmospheric turbulence-free case using the Gaussian beam is also plotted as a reference. From Figs. \ref{noncoded2x2}(a) and \ref{noncoded2x2}(b), we notice that as the topological charge $p$ increases the BER decreases  and the optimal performance is reached for  the set $\left ( -10, +10 \right )$. For weak atmospheric turbulence, (as depicted in  \mbox{Fig. \ref{noncoded2x2}(a)}), excepting the set $\left ( -1, +1 \right )$, all other sets of OAM modes  reached the same performance as the AT-free channel at the forward error correction (FEC) limit  of $3.8 \times 10^{-3}$.  These results, clearly show that the choice of the OAM modes based on the minimization of the MDL is an accurate criterion to obtain the lowest error probability performance. However, as can be seen from \mbox{Fig. \ref{noncoded2x2}(b)}, in the strong turbulence regime, the optimal OAM set $\left ( -10, +10 \right )$ could not completely compensate for AT.
\newline
In Fig. \ref{noncoded3x3}, we consider a $3 \times 3$ MIMO system. For both the weak and strong AT regimes, we notice that the lowest BER is achieved by the OAM set  $\left ( -10, 0, +10 \right )$. For the weak turbulence regime (as depicted in  \mbox{Fig. \ref{noncoded3x3}(a)}) the latter set of modes achieves the performance of the AT-free channel at BER below the FEC limit. For strong atmospheric turbulence (see \mbox{Fig. \ref{noncoded3x3}(b)}) the optimal set of OAM modes also achieves the lowest BER. Nonetheless, the performance degrades significantly due to the severity of the turbulence. To further enhance, the obtained performance, we propose to add a space-time coding scheme at the transmitter.
\section{Space-Time Coding}
Space-time coding was initially designed for MIMO wireless communication to bring coding gain and achieve a full-diversity at the transmitter side in a constantly varying channel. Recently, ST coding was also investigated for optical communication and was demonstrated to be efficient in mitigating non-unitary effects in polarization multiplexed systems \cite{Awwad} and also in few-mode fibers \cite{amhoud,Awwad_JLT}. The ST coding principle consists in transmitting  a coded linear combination of modulated signals during several channel uses. At the receiver side, the ML detector estimates the data from the different copies  present on all modes which allows giving a better estimate. Different ST code families have been  designed for wireless MIMO channels such as ST block codes (STBC) \cite{naguib1998applications}, and ST trellis codes (STTC) \cite{tarokh}. We particularly focus here on the STBC.
\newline
For a $2 \times 2$ MIMO channel, and during two channel uses, a ST codeword matrix uses 4 modulated symbols to achieve a full-rate transmission. In our analysis, we use the Golden code \cite{Belfiore} and the Silver code \cite{tirkkonen} which are known to be the two best ST codes satisfying a full-rate, full-diversity and optimal coding gain.
In addition to a full-rate and full-diversity, the Golden code \cite{Belfiore} achieves the best coding gain for Rayleigh fading channels given by its minimum determinant equal to $\frac{1}{5}$. The design of the Golden code is based on the Golden number $\frac{1+\sqrt{5}}{2}$ used for the code construction. The codeword matrix is given by:
\begin{equation}
\mathbf{X}_{\text{Golden}}=\frac{1}{\sqrt{5}}\begin{bmatrix}
\alpha (s_1+\theta s_2) & \alpha (s_3 + \theta s_4)\\ 
i\bar{\alpha} (s_3+\bar{\theta} s_4) & \bar{\alpha}(s_1+\bar{\theta} s_2)
\end{bmatrix},
\end{equation}
where $\theta=\frac{1+\sqrt{5}}{2}$, $\bar{\theta}=\frac{1-\sqrt{5}}{2}$, $\alpha = 1+i+ i \theta$, $\bar{\alpha}=1+i+ i \bar{\theta}$, and $\left \{ s_1, s_2, s_3 ,s_4 \right \}$ are the modulated QPSK symbols.
\newline
The Silver code has also a full-rate and full-diversity for $2 \times 2$ MIMO Rayleigh fading channels. Its minimum determinant is equal to $\frac{1}{7}$, which explains the slightly lower performance than the Golden code.  However, the Silver code has a reduced decoding complexity~\cite{biglieri2009fast}. 
The Silver codeword matrix is given by:
\begin{align}
\mathbf{X}_{\text{Silver}}&=\mathbf{X}_{\mathit{1}}(s_1,s_2)+\mathbf{TX}_{\mathit{1}}(z_1,z_2)\\
&=\begin{bmatrix}
s_1 & -s_2^* \\ 
s_2 & s_1^*
\end{bmatrix}+\begin{bmatrix}
1 &0 \\ 
0 & -1
\end{bmatrix}\begin{bmatrix}
z_1 & -z_2^* \\ 
z_2 & z_1^*
\end{bmatrix},
\end{align}
with $z_1$ and $z_2$ are given by:
\begin{equation*}
\begin{bmatrix}
z_1\\ 
z_2
\end{bmatrix}=\frac{1}{\sqrt{7}}\begin{bmatrix}
1+i &-1+2i \\ 
1+2i & 1-i
\end{bmatrix}\begin{bmatrix}
s_3\\ 
s_4
\end{bmatrix}.
\end{equation*}

\begin{figure*}[!h]
	\centering
	\begin{subfigure}[b]{0.5\textwidth} 
		\includegraphics[width=\columnwidth,height=7.5cm]{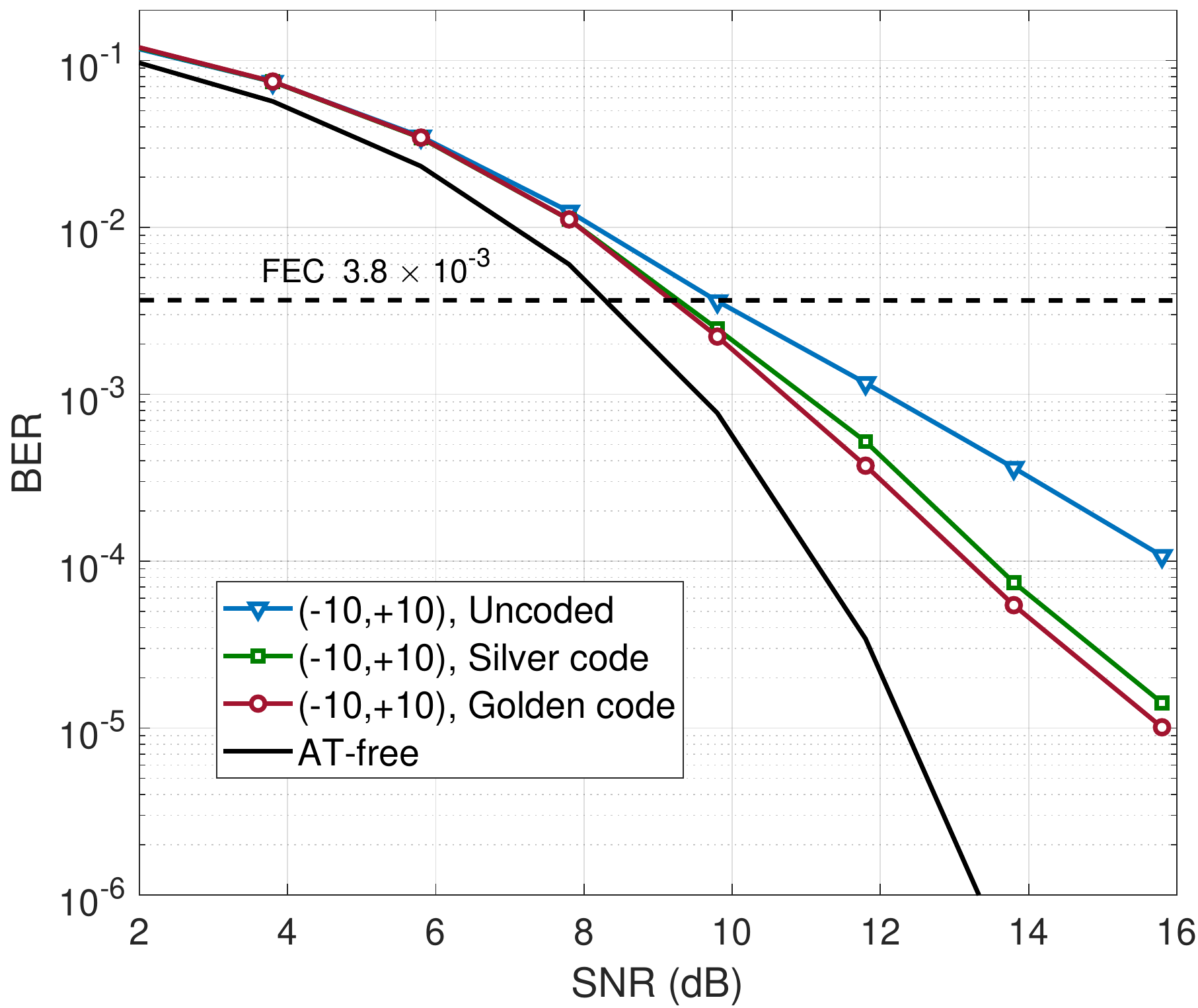}
		\caption{\footnotesize{ $2 \times 2 $ MIMO system in the strong turbulence regime, $C^2_n=10^{-14}$. } }
	\end{subfigure}~
	\begin{subfigure}[b]{0.5\textwidth} 
		\includegraphics[width=\columnwidth,height=7.5cm]{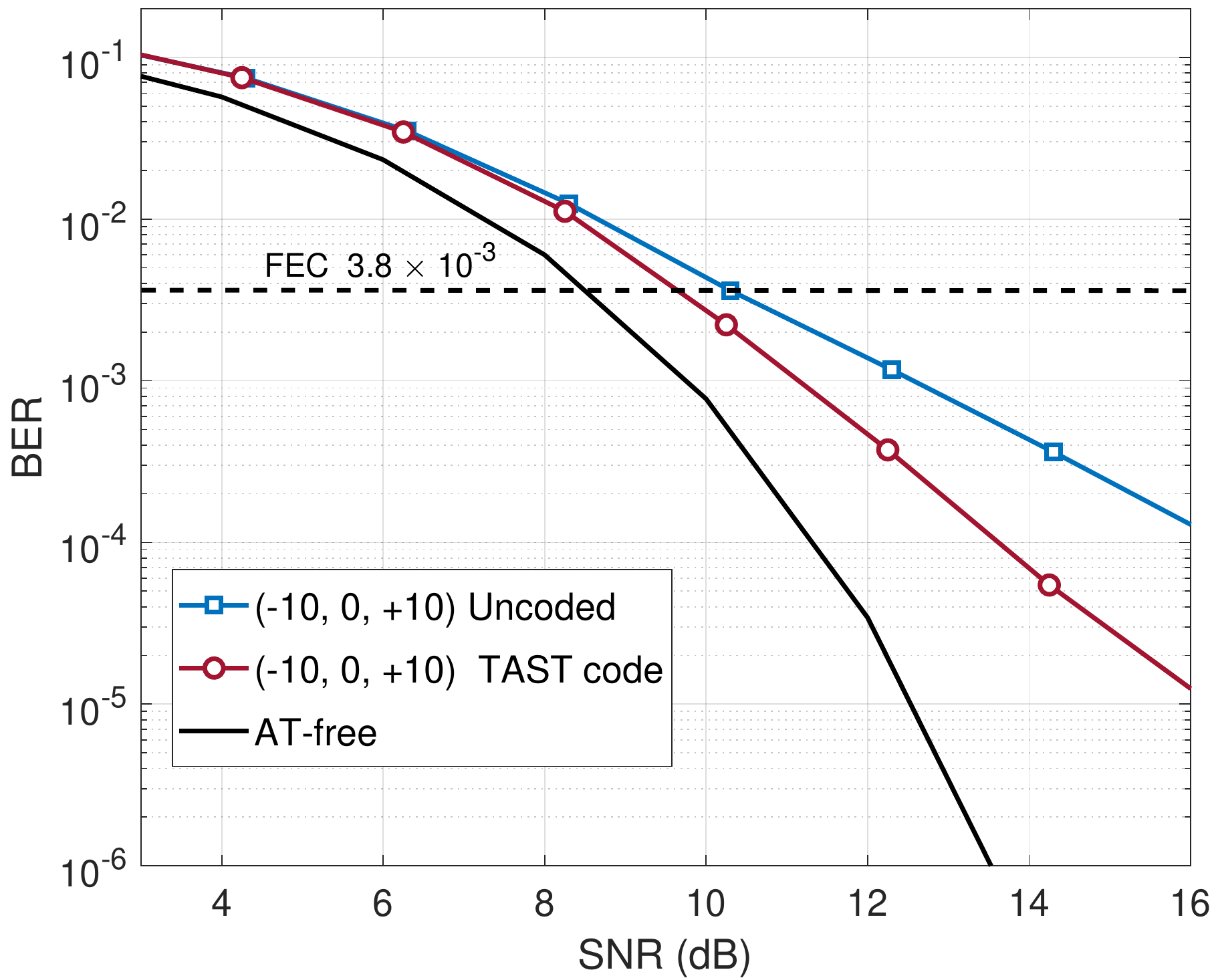}
		\caption{\footnotesize{$3 \times 3 $ MIMO system in the strong turbulence regime, $C^2_n=10^{-13}$. } }
	\end{subfigure}
	\caption{BER versus SNR for different OAM  sets for a $2 \times 2 $ and $3 \times 3 $ MIMO systems.}
	\label{coded}
\end{figure*}

Furthermore, for the $3 \times 3$ MIMO system, we consider a threaded algebraic ST code (TAST) \cite{TAST}.  The main advantage of the TAST code family is that it can achieve a full-rate and full-diversity for any $M \times M$ MIMO scheme \cite{TAST}. The TAST codeword matrix for a $3 \times 3$ MIMO channel is given by Eq. (\ref{TAST3x3}).
\begin{figure*}[b] \rule{18cm}{0.5pt} \\
	\begin{equation}\label{TAST3x3}  
	\mathbf{X}_{\text{TAST}_{3\times 3}}=   \frac{1}{\sqrt{3}}
	\begin{bmatrix}
	s_{1}+\theta s_{2}+\theta^{2}s_{3} & \phi^{2/3}(s_{7}+j\theta s_{8}+j^{2}\theta^{2}s_{9}) & \phi^{1/3}(s_{4}+j^{2}\theta s_{5}+j\theta^{2}s_{6}) \\
	\phi^{1/3}(s_{4}+\theta s_{5}+\theta^{2}s_{6}) & s_{1}+j\theta s_{2}+j^{2}\theta^{2}s_{3} & \phi^{2/3}(s_{7}+j^{2}\theta s_{8}+j\theta^{2}s_{9}) \\
	\phi^{2/3}(s_{7}+\theta s_{8}+\theta^{2}s_{9}) & \phi^{1/3}(s_{4}+j\theta s_{5}+j^{2}\theta^{2}s_{6}) & s_{1}+j^{2}\theta s_{2}+j\theta^{2}s_{3}
	\end{bmatrix}\\
	\end{equation}
	\centering{where $\phi=\exp(i\pi/12)$, $j=\exp(i2\pi/3) $, $\theta=\exp(i\pi/9)$ and $s_{i=1:9}$ are QPSK symbols.}
	
\end{figure*}
\newline
To have an insight on the performance of ST coding on OAM FSO transmission over a turbulent channel, we compare ST coded transmissions using  the Golden and Silver codes to  the uncoded transmission. A QPSK constellation is used to construct the codewords of the ST codes. We measure the BER performance using Monte-Carlo simulations, and for each
simulated point, we record a minimum of 100 bit errors. In Fig.~\ref{coded}(a), we plot the BER curves versus the SNR for the strong AT regime for a $ 2 \times 2$ MIMO system using the OAM set $\left ( -10, +10 \right )$. From the figure, we notice that the ST coded scheme provides an important coding gain compared to the uncoded transmission. The Golden code outperforms the Silver code, and  the coding gain obtained at BER=$10^{-4}$ by the Golden code is 2.2 dB  compared to the uncoded scheme. Hence the SNR gap to  the turbulence-free channel is reduced to 2.6 dB. In Fig.~\ref{coded}(b), we consider a $ 3 \times 3$ MIMO system using the OAM set $\left ( -10, 0, +10 \right )$ in the strong atmospheric turbulence regime. From the figure, we notice that at a BER=$10^{-4}$ the TAST code provides 2.2 dB gain compared to the uncoded scheme. Consequently, the obtained results, show that ST coding is an efficient DSP coding technique capable of mitigating atmospheric turbulence in OAM FSO systems.

\section{Conclusion}
In summary, we have shown that an optimal selection of OAM modes is relevant to improve the performance of OAM
FSO systems over the turbulent atmosphere. The selection criterion is based on the minimization of the MDL without any algorithm  required to update the set of optimal OAM modes. Hence, it is a low-cost complexity solution that can be integrated into real-time systems. In our simulations, we have considered an ML decoding strategy to obtain optimal performances. Nonetheless, sub-optimal decoders with lower complexity can  be also used. To further mitigate the atmospheric turbulence effect on FSO transmission, we proposed a ST coding scheme at the transmitter. We showed that AT was completely mitigated in the weak turbulence regime and important coding gains were obtained in the strong turbulence regime. Future work will extend to experimental validation of the proposed techniques. The use of OAM modes with non-null radial index will be equally considered.

\printbibliography

\EOD
\end{document}